\documentclass{article}

\usepackage{arxiv}
\usepackage[utf8]{inputenc}
\usepackage[T1]{fontenc}
\usepackage{hyperref}
\usepackage{url}
\usepackage{booktabs}
\usepackage{amsmath,amssymb}
\usepackage{graphicx}
\usepackage[numbers,sort&compress]{natbib}

\newcommand{\CodeUrl}{https://github.com/wizrads/MouseSeg}
\newcommand{\WebUrl}{https://mouseseg.joeyschulz.com}

\newcommand{\NPublicTotal}{677}
\newcommand{\NJensenTotal}{452}
\newcommand{\NRosenhainTotal}{225}
\newcommand{\NTrainScans}{455}
\newcommand{\NValScans}{98}
\newcommand{\NPublicTest}{124}

\newcommand{\NInstMicroCT}{9}
\newcommand{\NInstLowDoseCT}{48}
\newcommand{\NInstLowDoseMice}{12}
\newcommand{\NTotalUnseen}{181}

\newcommand{\UniParams}{526{,}947}
\newcommand{\UniFPSize}{2.03}
\newcommand{\UniIntSize}{0.96}
\newcommand{\SpecParams}{522{,}012}
\newcommand{\SpecFPSize}{2.01}
\newcommand{\SpecIntSize}{0.94}

\newcommand{\HumanTenOARDice}{0.6868}

\newcommand{\UniLatencyMThree}{1.56}
\newcommand{\UniTestLatencyPC}{9.30}
\newcommand{\UniTestOARDice}{0.8366}
\newcommand{\UniTestTumDice}{0.8522}
\newcommand{\UniTestTumMed}{0.9260}
\newcommand{\UniTestTumHD}{1.27}
\newcommand{\UniTestTumHDMed}{0.51}

\newcommand{\UniTestVolR}{0.9762}

\newcommand{\UniTestSens}{97.1\%}
\newcommand{\UniTestSpec}{100\%}
\newcommand{\UniTestSensSpec}{97.1\% / 100\%}
\newcommand{\UniTestLesionSens}{95.9\%}
\newcommand{\UniTestLesionPrec}{98.3\%}
\newcommand{\UniTestJenInDomDice}{0.8564}
\newcommand{\UniTestJenInDomMed}{0.9244}
\newcommand{\UniTestJenInDomHD}{1.35}
\newcommand{\UniTestDOneNineDice}{0.9259}
\newcommand{\UniTestDOneNineMed}{0.9534}
\newcommand{\UniTestDOneNineHD}{0.68}
\newcommand{\UniTestPriorSwinUnseenDice}{0.8908}
\newcommand{\UniTestPriorSwinUnseenMed}{0.9148}
\newcommand{\UniTestPriorSwinUnseenHD}{0.85}
\newcommand{\UniTestRosenContrastDice}{0.6578}
\newcommand{\UniTestRosenContrastMed}{0.7743}
\newcommand{\UniTestRosenContrastHD}{2.25}
\newcommand{\UniTestMultiRaterStaple}{0.8778}

\newcommand{\SpecAllTumDiceMean}{0.8519}
\newcommand{\SpecAllTumDiceMed}{0.9268}
\newcommand{\SpecAllTumHDMean}{1.18}
\newcommand{\SpecAllTumHDMed}{0.59}

\newcommand{\SpecAllSens}{96.1\%}
\newcommand{\SpecAllSpec}{100\%}

\newcommand{\SpecPCLatency}{5.90}

\newcommand{\LowDoseUniDice}{0.7501}
\newcommand{\LowDoseUniDiceMed}{0.9019}
\newcommand{\LowDoseUniHD}{3.31}

\newcommand{\LowDoseUniLat}{6.29}
\newcommand{\LowDoseSpecDice}{0.8481}
\newcommand{\LowDoseSpecDiceMed}{0.9102}
\newcommand{\LowDoseSpecHD}{2.11}

\newcommand{\LowDoseSpecLat}{2.50}

\newcommand{\CampUniAISegDice}{0.8975}
\newcommand{\CampUniAISegDiceMed}{0.9232}
\newcommand{\CampUniAISegHD}{2.58}

\newcommand{\CampUniLat}{2.62}

\newcommand{\CampSpecAISegDice}{0.9137}
\newcommand{\CampSpecAISegDiceMed}{0.9247}
\newcommand{\CampSpecAISegHD}{0.98}
\newcommand{\CampSpecManualDice}{0.7400}
\newcommand{\CampSpecManualDiceMed}{0.7656}
\newcommand{\CampSpecManualHD}{5.78}
\newcommand{\CampSpecLat}{2.07}

\newcommand{\ParityMeanDice}{0.9887}

\newcommand{\UniParityTumDice}{0.9853}

\newcommand{\AirFixLowDosePct}{1.09\%}

\title{MouseSeg: Open Source 3D Mouse Organ and Tumor CT Segmentation in Under 10 Seconds Without GPUs}

\author{
  Joseph B. Schulz, MS$^{1}$\thanks{Corresponding author: \texttt{jschulz9@wisc.edu}} \quad
  Campbell Haasch, MS$^{1}$ \quad
  Ohyun Kwon, PhD$^{1,2}$ \quad
  Bryan P. Bednarz, PhD$^{1}$ \\[4pt]
  \multicolumn{1}{@{}c@{}}{\small $^{1}$Department of Medical Physics, School of Medicine and Public Health, University of Wisconsin, Madison, WI 53705, USA} \\
  \multicolumn{1}{@{}c@{}}{\small $^{2}$Department of Radiation Medicine, University of Wisconsin, Madison, WI 53792, USA}
}

\renewcommand{\shorttitle}{MouseSeg: Open Source Mouse CT Segmentation Without GPUs}

\hypersetup{
  hidelinks,
  pdftitle={MouseSeg: Open Source 3D Mouse Organ and Tumor CT Segmentation in Under 10 Seconds Without GPUs},
  pdfauthor={Joseph B. Schulz, Campbell Haasch, Ohyun Kwon, Bryan P. Bednarz},
  pdfsubject={Preclinical murine microCT and cone beam CT 3D organ and subcutaneous tumor segmentation on CPU and WebAssembly},
  pdfkeywords={MouseSeg, mouse microCT segmentation, murine CT organ segmentation, subcutaneous xenograft tumor segmentation, preclinical radiotherapy, small animal cone beam CT, CBCT, organs at risk, radiopharmaceutical dosimetry, 3D U-Net, ONNX Runtime, CPU segmentation, WebAssembly, NIfTI}
}
\newcommand{\doi}[1]{doi:\,\href{https://doi.org/#1}{\nolinkurl{#1}}}

\begin{document}
\maketitle

\begin{abstract}
\textbf{Background:} Preclinical radiotherapy, dosimetry, and oncology studies require 3D contours of murine organs at risk and subcutaneous xenograft tumors in microCT and cone beam computed tomography (CT). Existing models rely on large graphics processing unit (GPU) weights and often fail on noisy low dose CT or clipped air values.
\textbf{Purpose:} To develop, distill, quantize, and validate MouseSeg, an open source CPU tool under $1$~MB for 10 mouse organs and subcutaneous tumors with NIfTI preprocessing, cradle stripped body masks, Python, and browser execution.
\textbf{Methods:} A unified dual window 3D U-Net (\UniParams{} parameters, \UniIntSize{}~MB) and a 3D tumor specialist (\SpecParams{} parameters, \SpecIntSize{}~MB) were distilled from SwinUNETR teachers across \NPublicTotal{} public scans. Preprocessing repaired NIfTI headers, restored clipped air values, resampled to $0.20$~mm, standardized soft tissue Hounsfield units (HU), and stripped cradles. Evaluation covered \NTotalUnseen{} unseen scans across public test sets, institutional microCT, and $1$~mA flat panel cone beam CT.
\textbf{Results:} On \NPublicTest{} held out public scans, the unified 3D network reached \UniTestOARDice{} organ Dice, \UniTestTumDice{} tumor Dice, and \UniTestSpec{} healthy specificity in \UniTestLatencyPC{}~s on a desktop CPU. Across institutional $1$~mA cone beam CT and microCT scans, both models detected every tumor with zero isolated false positives while matching full precision accuracy.
\textbf{Conclusions:} Standardized NIfTI preprocessing and knowledge distillation enable fast 3D murine organ and tumor segmentation under $1$~MB on standard CPUs and browsers. The browser app is at \url{\WebUrl} and Python code is at \url{\CodeUrl}.
\end{abstract}


\section{Introduction}

Preclinical radiation biology, radiopharmaceutical dosimetry, and oncology studies depend on 3D contours of tumors and organs at risk in CT. Image guided small animal irradiators plan treatments on onboard cone beam CT \cite{Verhaegen2011,Wong2008}, radiopharmaceutical therapy studies need organ masks for absorbed dose estimates \cite{Sgouros2020,Bolch2009}, and xenograft studies track tumor growth by microCT volume \cite{JensenMM2008}. Manual contouring across serial microCT scans is slow and shows rater variability along soft tissue borders \cite{Rosenhain2018,Jensen2024,Ali2013}. Although human CT tools such as TotalSegmentator \cite{Wasserthal2023} and MOOSE \cite{ShiyamSundar2022} standardized clinical segmentation, preclinical workflows still depend on interactive vendor software \cite{Gremse2016}, atlas registration \cite{Baiker2010,vanderHeyden2019}, single site scripts, or GPU deep learning models \cite{Schoppe2020,Malimban2022,Lappas2022,Jiang2024,Jensen2026}. Published preclinical networks include AIMOS \cite{Schoppe2020}, nnU-Net \cite{Isensee2021} thorax models \cite{Malimban2022}, AutoMICE built on SwinUNETR \cite{Jiang2024,Hatamizadeh2022}, and the TumSeg 3D U-Net \cite{Jensen2026}. In local CPU tests, their public weights reached $141$ to $187$~MB, an in house SwinUNETR tumor model reached $719$~MB, and AutoMICE inference took up to $35.7$~minutes on a laptop CPU. Models trained only on clean microCT scans often fail on noisy low dose cone beam CT with streak, ring, and cupping artifacts \cite{Schulze2011}, clipped air backgrounds, or broken NIfTI headers. Separate organ and tumor models can also create overlapping contours inside healthy organs. An open preclinical CT tool modeled after MOOSE and TotalSegmentator should run without GPUs, tolerate cone beam noise, and prevent organ and tumor overlap. It should also remove imaging cradles for dose calculation and run identically in Python and local web browsers. The purpose of this study was to develop, distill, quantize, and validate such a 3D CT suite across \NTotalUnseen{} unseen scans.

\section{Methods}

\subsection{Datasets and Evaluation Splits}

Training and testing combined two public microCT datasets with two institutional CT cohorts across \NTotalUnseen{} unseen test scans. Whole body organ masks came from Rosenhain et al. \cite{Rosenhain2018}, covering \NRosenhainTotal{} scans from $30$ mice. The default segmentation set covers $10$ organs: urinary bladder, lung, heart, liver, intestine, kidney, spleen, stomach, trachea, and bone. Two Rosenhain labels were excluded. Brain is absent from the source atlas, and the thigh muscle label is a small $5$ to $72$~mm$^3$ tracer reference patch without clear anatomical borders in CT. Subcutaneous tumor masks came from Jensen et al. \cite{Jensen2024}, covering \NJensenTotal{} scans from $223$ mice, each contoured by three annotators. Three rater consensus masks were formed with STAPLE \cite{Warfield2004}, and the $91$ Jensen test scans were used for human rater comparisons.

The \NPublicTotal{} public scans were split by mouse into training, validation, and held out test sets (\NTrainScans{}, \NValScans{}, and \NPublicTest{} scans). The public test set included healthy controls, contrast scans with both organs and tumors, in domain tumor scans, and external Dataset 9 scans. The tumor teacher (PriorSwin) was a previously trained in house SwinUNETR \cite{Hatamizadeh2022} tumor model. No test mouse appeared in the training splits, in the published AutoMICE training set \cite{Jiang2024}, or in the PriorSwin training set. Independent external testing used two CT scanners at the University of Wisconsin. The first institutional set contained \NInstMicroCT{} high resolution microCT scans with 4T1, EBC1, and CHLA20 tumors alongside in house SwinUNETR and manual contours. The second set contained \NInstLowDoseCT{} serial flat panel CT scans acquired at $1$~mA without contrast from \NInstLowDoseMice{} xenograft mice alongside in house SwinUNETR tumor masks.

\subsection{CT Preprocessing and Coordinate Standardization}

Each input NIfTI scan underwent automated spatial and intensity preprocessing before inference. Headers were converted to right anterior superior (RAS) orientation, and broken unit sform headers were repaired using true qform voxel spacing. When orientation tags were uncertain, an automated intensity search selected the right handed axis order matching lung and bone profiles. Workstation exports often clip outside air at $-400$~HU, which distorts resampling near the skin and inflates body masks. A histogram check before resampling detected clipped air backgrounds and restored outside air voxels to $-1000$~HU while leaving normal scans unchanged. Volumes were resampled to $0.20$~mm isotropic voxels using smoothed downsampling and trilinear interpolation. Before inference, a slice consistency check removed stationary scanner rails and rods from the bounding box without changing HU values inside the mouse. To reduce cone beam calibration drift, soft tissue intensities were shifted smoothly toward $+45$~HU, and broad ($[-1000, 3000]$~HU) plus soft tissue ($[-400, 1000]$~HU) windows were scaled to $[0, 1]$. The tumor specialist received only the soft tissue window inside a tight bounding box around the mouse.

\subsection{Compact 3D Network, Noise Simulation, and Distillation}

\begin{figure}[tbp]
  \centering
  \includegraphics{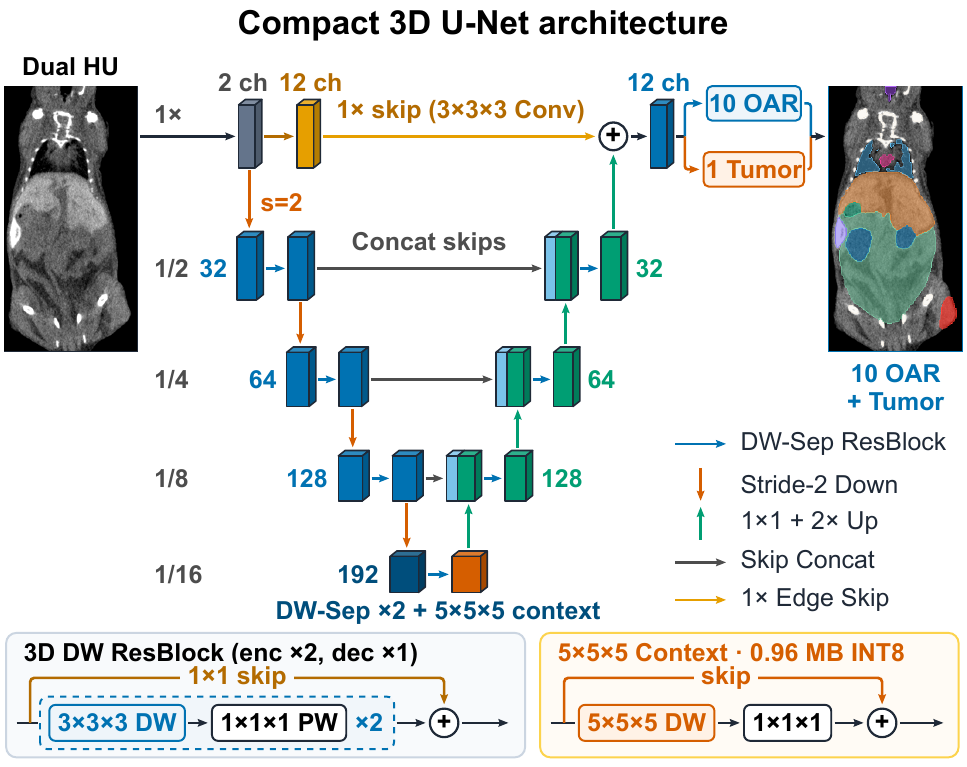}
  \caption{Compact 3D U-Net architecture (\UniParams{} parameters, \UniIntSize{}~MB INT8). Dual window $0.20$~mm CT inputs (coronal mouse CT) pass through a stride $2$ encoder ($32$ to $192$ channels), $5\times5\times5$ bottleneck block, trilinear decoder with skip connections, and full resolution residual addition before the $10$ organ and $1$ tumor output heads.}
  \label{fig:arch}
\end{figure}

Both models used a compact 3D U-Net style encoder decoder network \cite{Ronneberger2015,Cicek2016} built for fast CPU inference (\textbf{Fig.~\ref{fig:arch}}). A stride $2$ input layer fed four depthwise separable \cite{Howard2017} $3\times3\times3$ residual \cite{He2016} stages with $(32, 64, 128, 192)$ channels, a $5\times5\times5$ bottleneck block, and trilinear upsampling. A full resolution skip path summed directly into the decoder output using $12$ channels in the unified model (\UniParams{} parameters) and $8$ channels in the tumor specialist (\SpecParams{} parameters), with auxiliary training heads removed before export. Online physics augmentation added realistic CT noise to training patches ($96\times160\times160$), including quantum noise, blur, lower resolution, ring artifacts, beam hardening, bone streaks, and contrast shifts typical of cone beam CT \cite{Schulze2011}. Networks trained on an NVIDIA RTX A6000 GPU using PyTorch \cite{Paszke2019}, MONAI \cite{Cardoso2022}, AdamW \cite{Loshchilov2019}, and mixed precision \cite{Micikevicius2018} for $45$ epochs (unified) or $37$ epochs (specialist), followed by batch normalization updates on real scans. Organ training combined masked cross entropy and Dice loss \cite{Milletari2016} with extra weight on organ borders, following boundary aware segmentation losses \cite{Kervadec2021}. Tumor training combined masked focal binary cross entropy \cite{Lin2017}, Dice loss \cite{Milletari2016}, and healthy tumor free controls. Knowledge distillation \cite{Hinton2015} transferred soft probabilities from AutoMICE \cite{Jiang2024} and the PriorSwin tumor teacher alongside Mean Teacher consistency \cite{Tarvainen2017}. External comparisons used the published AutoMICE \cite{Jiang2024} and TumSeg \cite{Jensen2026} models, the in house PriorSwin model, and human rater contours.

\subsection{Quantization, Postprocessing, Body Mask, and Browser Tool}

ONNX Runtime \cite{ONNXRuntime} models underwent static 8 bit integer (INT8) quantization \cite{Jacob2018} calibrated on representative CT scans. Input, depthwise 3D, full resolution skip, and output layers stayed in 32 bit floating point (FP32) precision while $1\times1\times1$ convolutions ran in INT8. Sliding window inference processed overlapping $128$ slice axial windows ($25\%$ overlap) across the full axial cross section so both flanks remained in the field of view. Tumor probabilities underwent 3D hysteresis thresholding calibrated on institutional $1$~mA CT (seed $0.78$ and grow $0.18$ for the unified model, seed $0.80$ and grow $0.40$ for the specialist) with soft tissue HU and minimum volume limits. Default inference excluded the auxiliary thigh muscle channel to output $10$ organs plus tumor, using a bladder threshold ($0.18$) calibrated on institutional scans along with minimum probability thresholds on lung, intestine, and trachea.

Connected component postprocessing in 3D anchored organ masks to the heart position before resolving organ and tumor overlap. Single organs (urinary bladder, heart, liver, intestine, spleen, stomach, and trachea) retained only the largest connected component. Paired lungs and kidneys retained at most two components on resampled and native grids, whereas bone and tumor masks allowed multiple components. Lung masks were trimmed inside the ribcage using bone borders and air HU thresholds. Trachea masks retained one air filled airway expanded slightly into the tracheal wall, and small intestine false positives near the head or bladder were removed using head to tail position limits and morphological opening. Small enclosed cavities inside organ masks were filled from adjacent labels, and internal organs were protected from tumor overlap. Morphological cradle removal then stripped animal holders and filled internal cavities to output a separate whole body mask enclosing all segmented structures. Python ran INT8 weights, whereas the static WebAssembly \cite{Haas2017} web app ran FP32 weights locally in the browser using single instruction multiple data (SIMD) instructions without uploading scans.

\subsection{Validation Metrics}

Accuracy was measured with the Dice similarity coefficient, 95th percentile Hausdorff distance ($\text{HD}_{95}$), average symmetric surface distance (ASSD), relative volume error, scan sensitivity, healthy specificity, and lesion matching by volume. Volume agreement used Pearson $r$ and Lin concordance correlation coefficient ($\text{CCC}$) \cite{Lin1989}. Human agreement was computed from $17$ held out Rosenhain test scans with paired second rater masks \cite{Rosenhain2018} and from the $91$ three rater Jensen test scans with STAPLE consensus \cite{Jensen2024,Warfield2004}. Runtime and peak memory were measured on an Apple M3 Pro laptop, a desktop PC, and two thread workstations, along with FP32 versus INT8 checks and $-400$~HU air restoration tests.

\section{Results}

\subsection{Whole Body Organ and Subcutaneous Tumor Benchmark Performance}

\begin{figure}[tbp]
  \centering
  \includegraphics{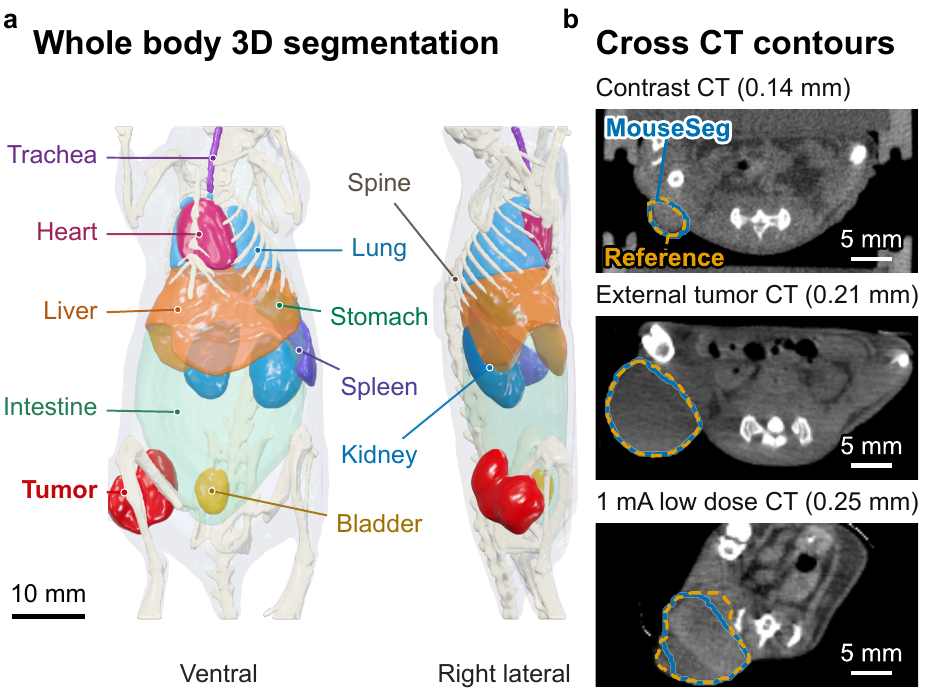}
  \caption{Three dimensional whole body segmentation and CT contour comparison across scanners. (a) Front and right side WebGL surface views of unified 3D predictions on a held out contrast mouse CT ($0.20$~mm grid) showing $10$ body organs, skeleton, cradle stripped body surface, and a $265.6$~mm$^3$ right flank tumor. (b) Axial tumor contours comparing solid blue unified 3D predictions with dashed orange reference masks on held out contrast CT ($0.14$~mm), external Dataset~9 ($0.21$~mm STAPLE), and institutional $1$~mA low dose CT ($0.25$~mm).}
  \label{fig1}
\end{figure}

\begin{figure}[tbp]
  \centering
  \includegraphics{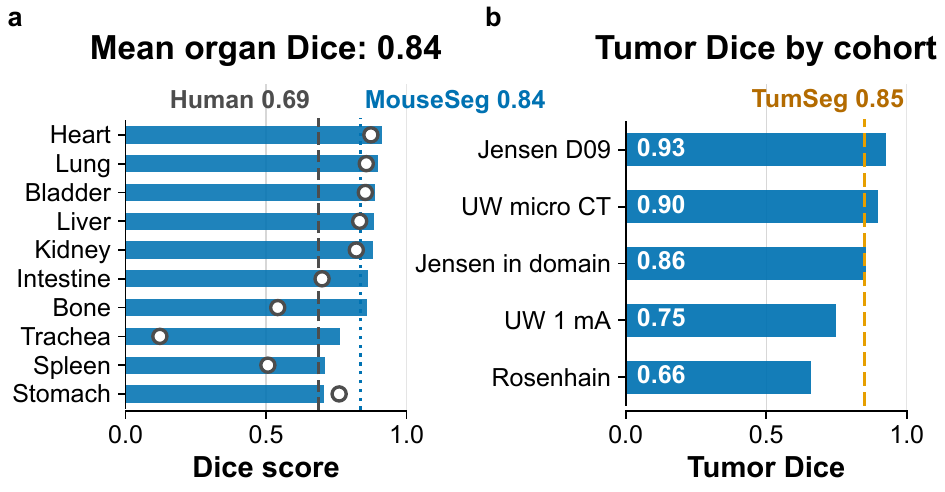}
  \caption{Organ and tumor segmentation accuracy across cohorts. (a) Mean 3D Dice across $10$ Rosenhain organs for the unified 3D model (blue bars, mean \UniTestOARDice{}) versus human rater agreement (gray circles, mean \HumanTenOARDice{}). (b) Mean unified 3D tumor Dice across five unseen public and institutional CT cohorts compared against TumSeg \cite{Jensen2026} run on the same held out Jensen scans ($0.8499$, dashed orange line).}
  \label{fig2}
\end{figure}

Across \NPublicTest{} held out public CT scans, the \UniIntSize{}~MB unified 3D network reached \UniTestOARDice{} mean organ Dice and \UniTestTumDice{} mean tumor Dice (\textbf{Table~\ref{tab:benchmark}}). Inference finished in \UniTestLatencyPC{}~s on a desktop PC, segmented whole body anatomy (\textbf{Fig.~\ref{fig1}a}), and matched manual tumor volumes closely ($r = \UniTestVolR{}$). Specificity on healthy controls reached \UniTestSpec{} with zero false positive volume, while scan sensitivity reached \UniTestSens{} by detecting every tumor above $50$~mm$^3$ and one quarter of smaller lesions. Across individual tumors, lesion sensitivity was \UniTestLesionSens{} and precision was \UniTestLesionPrec{}, with only two false positive components across all test scans. By organ, agreement ranged from $0.86$ to $0.91$ for heart, lung, urinary bladder, liver, kidney, intestine, and bone (\textbf{Fig.~\ref{fig2}a}), whereas trachea ($0.7640$), spleen, and stomach had lower overlap. Across all 10 organs, the unified model exceeded human rater agreement (\HumanTenOARDice{}), and across seven shared organs it outperformed AutoMICE ($0.8627$ versus $0.5787$). Anatomical postprocessing improved every organ over raw network output (raising 10 organ mean Dice from $0.8170$ to \UniTestOARDice{} at $0.88$~mm mean $\text{HD}_{95}$) with zero component count errors across $33$ scans.

\begin{table}[tbp]
  \centering
  \caption{Comparison of model size, CPU runtime, and accuracy against published preclinical models and human raters on \NPublicTest{} held out public CT scans. Dashes (-) indicate metrics that do not apply to a method or were not reported. Best model values in each column are shown in bold.}
  \label{tab:benchmark}
  \begingroup
  \normalsize
  \setlength{\tabcolsep}{3pt}
  \renewcommand{\arraystretch}{1.22}
  \begin{tabular*}{\textwidth}{@{\extracolsep{\fill}}llccccc@{}}
    \toprule
    Method &
    Task &
    \begin{tabular}[t]{@{}c@{}}Size\\(MB)\end{tabular} &
    \begin{tabular}[t]{@{}c@{}}CPU Time\\(s)\end{tabular} &
    \begin{tabular}[t]{@{}c@{}}Organ\\Dice\end{tabular} &
    \begin{tabular}[t]{@{}c@{}}Tumor\\Dice\end{tabular} &
    \begin{tabular}[t]{@{}c@{}}Tumor $\text{HD}_{95}$\\(mm)\end{tabular} \\
    \midrule
    \multicolumn{7}{@{}l}{\textit{Whole body organ segmentation}} \\
    Human raters ($n = 17$)$^{a}$ & 10 organs & - & Manual & \HumanTenOARDice{} & - & - \\
    AutoMICE \cite{Jiang2024}$^{b}$ & 7 organs & 141 & 224.5 (M3) & 0.5787$^{b}$ & - & - \\
    \midrule
    \multicolumn{7}{@{}l}{\textit{Subcutaneous tumor segmentation}} \\
    Human raters ($n = 91$)$^{a}$ & Tumor & - & Manual & - & 0.9110 to 0.9181 & - \\
    PriorSwin$^{c}$ & Tumor & 719 & GPU only & - & 0.7919 & 5.94 \\
    TumSeg \cite{Jensen2026}$^{d}$ & Tumor & 187 & 17.30 (M3) & - & 0.8499 & 2.24 \\
    \midrule
    \multicolumn{7}{@{}l}{\textit{Proposed CPU models (INT8)}} \\
    Unified 3D & 10 organs + tumor & \UniIntSize{} & \UniLatencyMThree{} / \UniTestLatencyPC{}$^{e}$ & \textbf{\UniTestOARDice{}} & \textbf{\UniTestTumDice{}} & \UniTestTumHD{} \\
    Tumor Specialist 3D & Tumor & \textbf{\SpecIntSize{}} & \textbf{1.06 / \SpecPCLatency{}$^{e}$} & - & \SpecAllTumDiceMean{} & \textbf{\SpecAllTumHDMean{}} \\
    \bottomrule
  \end{tabular*}
  \endgroup

  \vspace{6pt}
  \parbox{\textwidth}{\raggedright\small $^{a}$Interobserver agreement computed in this study: mean second rater Dice across 10 Rosenhain organs on $17$ held out test scans with paired masks \cite{Rosenhain2018} and pairwise Dice among three Jensen annotators on the $91$ test scans ($0.9522$ to $0.9597$ versus STAPLE). Jensen et al. report $0.900$ to $0.907$ pairwise Dice across all $452$ scans \cite{Jensen2024}. $^{b}$AutoMICE (35.1M parameters) evaluated on 7 overlapping organs across unseen Rosenhain mice on an 8 thread Apple M3 Pro CPU at $0.5$ overlap. $^{c}$In house SwinUNETR tumor model on 36 unseen Jensen scans. $^{d}$Released CT three model ensemble (49.0M parameters) on 65 clean held out Jensen scans (median Dice $0.9185$, median $\text{HD}_{95}$ $0.47$~mm, single AC model $62.3$~MB, $7.98$~s at $0.42$~mm). $^{e}$CPU time shown as 8 thread Apple M3 Pro / 4 thread desktop PC median across \NPublicTest{} scans (M3 range $1.56$ to $2.03$~s unified, $1.06$ to $1.36$~s specialist). Median tumor Dice and $\text{HD}_{95}$ were \UniTestTumMed{} and \UniTestTumHDMed{}~mm for Unified 3D (\UniTestSensSpec{} sens/spec) and \SpecAllTumDiceMed{} and \SpecAllTumHDMed{}~mm for Tumor Specialist 3D (\SpecAllSens{} / \SpecAllSpec{} sens/spec).}
\end{table}

Accuracy stayed consistent across held out test subsets (\textbf{Fig.~\ref{fig1}b}, \textbf{Fig.~\ref{fig2}b}, \textbf{Table~\ref{tab:cohorts}}). On Dataset 9 and scans unseen by PriorSwin, the unified model matched or exceeded the GPU baselines with full tumor detection (\textbf{Table~\ref{tab:cohorts}}). Against three rater STAPLE consensus masks across $91$ scans, unified predictions scored \UniTestMultiRaterStaple{}, close to pairwise human agreement (\textbf{Table~\ref{tab:benchmark}}). The \SpecIntSize{}~MB tumor specialist reached similar overall tumor Dice in fewer seconds on the desktop CPU (\textbf{Table~\ref{tab:benchmark}}).

\subsection{External Validation on Institutional MicroCT and 1 mA Low Dose CT}

\textbf{Table~\ref{tab:cohorts}} summarizes results across all \NTotalUnseen{} unseen scans. On \NInstLowDoseCT{} flat panel CT scans acquired at $1$~mA with $50$ to $92$~HU noise, both models detected every tumor with zero isolated false positives. Under heavy noise, the 3D tumor specialist kept tighter tumor boundaries than the unified model (\textbf{Table~\ref{tab:cohorts}}), and specialist volumes tracked serial SwinUNETR growth curves ($r = 0.8179$). On \NInstMicroCT{} high resolution institutional microCT scans, both models also detected every tumor with zero isolated false positives (\textbf{Fig.~\ref{fig2}b}, \textbf{Table~\ref{tab:cohorts}}). Predictions closely matched SwinUNETR references, whereas manual contour Dice was lower because one manual contour included $11.19\%$ outside air beyond the skin. Across $101$ resampled validation scans and native $1$~mA confirmation, calibrated tumor hysteresis raised manual microCT Dice to $0.7666$ and $1$~mA Dice to $0.8312$ (unified) and $0.8380$ (specialist) with zero healthy false positives.

\begin{table}[tbp]
  \centering
  \caption{Evaluation across all \NTotalUnseen{} unseen public and institutional CT scans. Dashes (-) indicate metrics that do not apply to a subset. Best values in paired institutional comparisons are shown in bold.}
  \label{tab:cohorts}
  \begingroup
  \normalsize
  \setlength{\tabcolsep}{1.5pt}
  \renewcommand{\arraystretch}{1.24}
  \begin{tabular*}{\textwidth}{@{\extracolsep{\fill}}lcccccc@{}}
    \toprule
    Cohort / Test Subset &
    \begin{tabular}[t]{@{}c@{}}Scans\\($n$)\end{tabular} &
    Model Mode &
    \begin{tabular}[t]{@{}c@{}}Dice\\(Mean / Med)\end{tabular} &
    \begin{tabular}[t]{@{}c@{}}$\text{HD}_{95}$\\(mm)\end{tabular} &
    \begin{tabular}[t]{@{}c@{}}Sens /\\Spec\end{tabular} &
    \begin{tabular}[t]{@{}c@{}}Time\\(s)\end{tabular} \\
    \midrule
    Rosenhain Healthy & 21 & Unified 3D & - & - & - / \UniTestSpec{} & \UniTestLatencyPC{} (4T PC) \\
    Rosenhain Contrast & 12 & Unified 3D & \UniTestRosenContrastDice{} / \UniTestRosenContrastMed{} & \UniTestRosenContrastHD{} & 100\% / - & \UniTestLatencyPC{} (4T PC) \\
    Jensen In Domain & 63 & Unified 3D & \UniTestJenInDomDice{} / \UniTestJenInDomMed{} & \UniTestJenInDomHD{} & 95.2\% / - & \UniTestLatencyPC{} (4T PC) \\
    Jensen External D09 & 28 & Unified 3D & \UniTestDOneNineDice{} / \UniTestDOneNineMed{} & \UniTestDOneNineHD{} & 100\% / - & \UniTestLatencyPC{} (4T PC) \\
    PriorSwin Unseen & 31 & Unified 3D & \UniTestPriorSwinUnseenDice{} / \UniTestPriorSwinUnseenMed{} & \UniTestPriorSwinUnseenHD{} & 100\% / - & \UniTestLatencyPC{} (4T PC) \\
    \midrule
    UW MicroCT (SwinUNETR) & 9 & Unified 3D & \CampUniAISegDice{} / \CampUniAISegDiceMed{} & \CampUniAISegHD{} & \textbf{100\%} / - & \CampUniLat{} (2T CPU) \\
    UW MicroCT (SwinUNETR) & 9 & 3D Specialist & \textbf{\CampSpecAISegDice{}} / \textbf{\CampSpecAISegDiceMed{}} & \textbf{\CampSpecAISegHD{}} & \textbf{100\%} / - & \textbf{\CampSpecLat{}} (2T CPU) \\
    UW MicroCT (Manual) & 9 & 3D Specialist & \CampSpecManualDice{} / \CampSpecManualDiceMed{} & \CampSpecManualHD{} & 100\% / - & \CampSpecLat{} (2T CPU) \\
    UW Low Dose (1 mA) & 48 & Unified 3D & \LowDoseUniDice{} / \LowDoseUniDiceMed{} & \LowDoseUniHD{} & \textbf{100\%} / - & \LowDoseUniLat{} (2T CPU) \\
    UW Low Dose (1 mA) & 48 & 3D Specialist & \textbf{\LowDoseSpecDice{}} / \textbf{\LowDoseSpecDiceMed{}} & \textbf{\LowDoseSpecHD{}} & \textbf{100\%} / - & \textbf{\LowDoseSpecLat{}} (2T CPU) \\
    \bottomrule
  \end{tabular*}
  \endgroup
\end{table}

\subsection{Runtime, Model Size, Quantization, and Air Clipping}

\begin{figure}[tbp]
  \centering
  \includegraphics{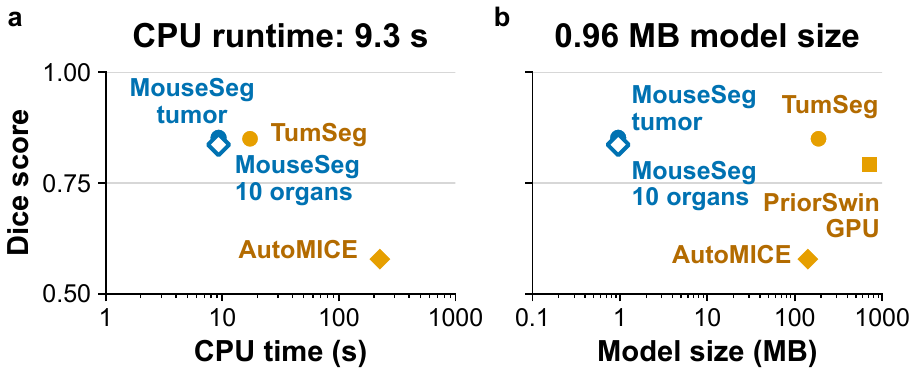}
  \caption{Accuracy versus runtime and model size against published preclinical baselines. (a) Mean Dice score versus CPU runtime on a log axis comparing unified 3D MouseSeg tumor (solid blue circle) and $10$ organ (open blue diamond) inference (\UniTestLatencyPC{}~s, four thread PC) against TumSeg ($17.30$~s) and AutoMICE ($224.5$~s, eight thread Apple M3 Pro). (b) Mean Dice score versus model size on a log axis comparing INT8 MouseSeg (\UniIntSize{}~MB) against AutoMICE ($141$~MB), TumSeg ($187$~MB), and GPU only PriorSwin ($719$~MB).}
  \label{fig3}
\end{figure}

INT8 quantization reduced the unified model from \UniFPSize{} to \UniIntSize{}~MB and the specialist from \SpecFPSize{} to \SpecIntSize{}~MB (\textbf{Fig.~\ref{fig3}b}), while both INT8 models closely matched FP32 tumor masks (\UniParityTumDice{} and \ParityMeanDice{} Dice). On an Apple M3 Pro CPU, whole volume inference took \UniLatencyMThree{} to $2.03$~s (unified) and $1.06$ to $1.36$~s (specialist). On the same CPU, the $141$~MB AutoMICE model took $224.5$~s to $35.7$~minutes (\textbf{Fig.~\ref{fig3}a}). Peak CPU memory stayed near $1$~GB on standard scans. Across $15$ scans, cropped 3D patches ran over twice as slowly as full cross section $128$ slice sliding windows without improving tumor Dice, whereas $25\%$ window overlap removed boundary artifacts for $0.26$~s of extra runtime. In the local browser tool without special server headers, single thread FP32 inference finished in $4.0$ to $15.6$~s across scans up to $395\times335\times819$ voxels. With ten thread cross origin isolation enabled, browser runtime dropped to $1.8$ to $5.2$~s and matched Python at $1.0000$ voxel agreement across $16$ runs. Across $62$ tested scans, restoring $-400$~HU clipped outside air before resampling prevented $+25.5\%$ to $+45.4\%$ body volume inflation along air borders and kept body volume error within \AirFixLowDosePct{} with zero clipped organ or tumor voxels.

\section{Discussion}

Compact 3D networks under $1$~MB distilled from SwinUNETR under simulated cone beam noise matched or exceeded large GPU models in seconds on standard CPUs and web browsers. Combining a stride $2$ encoder with a full resolution skip path and $5\times5\times5$ bottleneck preserved fine borders while rejecting ribs, cradles, and low dose noise. Protecting manual labels during distillation raised 10 organ Dice to \UniTestOARDice{} in one $0.96$~MB model and exceeded both AutoMICE and human rater agreement. Across external scanners, both models missed zero tumors and produced zero isolated false positives on $1$~mA flat panel CT and $0.08$~mm microCT near $1$~GB memory. The tumor specialist held tighter borders under heavy $1$~mA noise, whereas the unified model segmented all $10$ organs and the tumor in one pass. Restoring clipped air values before resampling also kept cradle stripped body masks within $1.1\%$, while Python and browser tools support local slice review, NIfTI mask export, and volume tables without server uploads. Following the TotalSegmentator \cite{Wasserthal2023} and MOOSE \cite{ShiyamSundar2022} model of one command, whole body output, these masks can feed small animal treatment planning \cite{Verhaegen2011} and organ level radiopharmaceutical dosimetry \cite{Bolch2009} without manual contouring.

This study has several limitations. First, organ contours cover the $10$ body structures in Rosenhain et al. \cite{Rosenhain2018} and omit brain, head and neck structures, and small tracer reference patches. Second, tumor testing covered subcutaneous flank tumors only and did not test internal or metastatic tumors. Third, hysteresis and minimum volume filtering kept $100\%$ healthy specificity yet missed three of four isolated tumors below $50$~mm$^3$ in the unified model ($50.0\%$ lesion detection under $50$~mm$^3$). Fourth, contrast tumor Dice and low contrast spleen or stomach overlap stayed lower than heart ($0.9131$) or lung ($0.8975$), and trachea postprocessing requires visible air below $-300$~HU inside the airway. Finally, browser WebAssembly runs FP32 weights faster than INT8 kernels in single thread SIMD whereas Python runs INT8 directly, and scanner specific fine tuning could further improve external microCT scans with shifted HU calibration.

\section{Conclusion}

Standardized CT preprocessing, noise augmentation, SwinUNETR distillation, and selective INT8 quantization enable 3D segmentation of 10 mouse organs and subcutaneous tumors in under $1$~MB without GPUs. Across \NTotalUnseen{} unseen scans, the unified model reaches \UniTestOARDice{} organ Dice, \UniTestTumDice{} tumor Dice, $100\%$ healthy specificity, zero isolated false positives on $1$~mA CT, and cradle stripped body masks across Python and local browser tools.

\section*{Acknowledgments and Availability}

The authors thank the developers and maintainers whose open source models, datasets, and libraries made this work possible, including Jiang et al. (AutoMICE) \cite{Jiang2024}, Jensen et al. (TumSeg and tumor database) \cite{Jensen2026,Jensen2024}, Rosenhain et al. \cite{Rosenhain2018}, Schoppe et al. (AIMOS) \cite{Schoppe2020}, Wasserthal et al. (TotalSegmentator) \cite{Wasserthal2023}, Shiyam Sundar et al. (MOOSE) \cite{ShiyamSundar2022}, Isensee et al. (nnU-Net) \cite{Isensee2021}, Hatamizadeh et al. (SwinUNETR) \cite{Hatamizadeh2022}, the MONAI Consortium \cite{Cardoso2022}, and the PyTorch \cite{Paszke2019} and ONNX Runtime \cite{ONNXRuntime} teams. Trained weights and Python command line tools are available at \url{\CodeUrl}, and the local browser WebAssembly app is available at \url{\WebUrl}, both released strictly for academic and noncommercial preclinical research use under the Creative Commons Attribution NonCommercial 4.0 International License (CC BY-NC 4.0). This release follows all upstream dataset, teacher, and library licenses (Rosenhain et al. CC0 1.0, Jensen et al. database on ERDA CC BY 4.0, TumSeg code CC BY-NC 4.0, AutoMICE Apache 2.0 and Zenodo CC BY 4.0, MONAI Apache 2.0, nnU-Net Apache 2.0, and ONNX Runtime Web MIT). The software and models are intended only for preclinical small animal research and are not cleared or approved for human clinical or diagnostic use.

\bibliographystyle{unsrtnat}
\bibliography{references}

\end{document}